\documentstyle[preprint,aps]{revtex}

\tighten
\preprint{FN-IEM 18}
\begin{document}
\draft
\title{Relativistic Analysis of the
$^{208}Pb(e,e'p)^{207}Tl$ reaction at High Momentum}

\author{J.M. Ud\'\i as\cite{tueb}, P. Sarriguren, E. Moya de Guerra,
and J.A. Caballero}
\address{
Instituto de Estructura de la Materia, \\ 
Consejo Superior de Investigaciones Cient\'\i ficas, \\ 
Serrano 119, E-28006 Madrid, Spain\\ 
}
\date{\today}
\maketitle
\begin{abstract}
The recent $^{208}$Pb$(e,e'p)^{207}$Tl data from NIKHEF-K at high 
missing momentum ($p_m>300$ MeV/c) are compared to theoretical 
results obtained with  a fully relativistic formalism previously 
applied to analyze data on the low missing momentum 
($p_m < 300$ MeV/c) region. The same relativistic optical potential 
and mean field wave functions are used in the two $p_m$-regions. 
The spectroscopic factors of the various shells are extracted from 
the analysis of the low-$p_m$ data and then used in the high-$p_m$ 
region. In contrast to previous analyses using a nonrelativistic 
mean field formalism, we do not find a substantial deviation from 
the mean field predictions other than that of the spectroscopic 
factors, which appear to be consistent with both low- and 
high-$p_m$ data. We find that the difference between results of 
relativistic and nonrelativistic formalisms is enhanced in the 
$p_m<0$ region that will be interesting to explore experimentally.

\end{abstract}


\maketitle

Coincidence $(e,e'p)$ measurements at quasielastic kinematics have 
been shown to provide very detailed information on the energy and
momentum distributions of the bound nucleons~\cite{Fru85}. This is 
so because at quasielastic kinematics the $(e,e'p)$ reaction can
be treated with confidence~\cite{Fru85} in the impulse approximation 
(IA), i.e., assuming that the detected knocked out proton absorbs 
the whole momentum ($q$) and energy ($\omega$) of the exchanged 
photon.

In the past $(e,e'p)$ experiments in parallel kinematics provided 
high precision measurements of reduced cross sections in the missing
momentum range $-50 < p_m < 300$ MeV/c~\cite{Quint}. This range of 
$p_m$-values was covered by varying the $q$-value while maintaining 
${\bf p}_m$ parallel to ${\bf q}$ (in what follows we refer to this 
as the low-$p_m$ region). Recently, the range of missing momentum 
has been extended by $(e,e'p)$ measurements at $(q,\omega)$-constant 
kinematics~\cite{Irene}.

The new range of $p_m$-values ($340 < p_m < 500$ MeV/c) was covered 
by varying the direction of the proton detector between $\sim 99^0$ 
and $\sim 140^0$, with fixed values of $q$ (221 MeV/c) and $\omega$ 
(110 MeV), at an incoming electron energy of 487 MeV. We will refer 
to the latter as the high-$p_m$ region. In both regions the kinetic 
energy of the detected proton was $T_F=100$ MeV.

In Ref.~\cite{Irene} the high-$p_m$ data for the shells $3s_{1/2}$, 
$2d_{3/2}$, $1h_{11/2}$, $2d_{5/2}$, and $1g_{7/2}$ in $^{208}$Pb, 
were compared with standard nonrelativistic calculations based on 
the {\sc dweepy} program developed in Ref.~\cite{Giusti}, that had 
also been used for the analysis of the low-$p_m$ data. The authors 
of Ref.~\cite{Irene} conclude that the high-$p_m$ data are 
substantially larger than the mean field predictions. The purpose 
of this paper is to see whether this conclusion still holds when 
the data are analyzed with the fully relativistic formalism 
recently developed~\cite{spect,poten}.

The simplest approximation to analyze the $(e,e'p)$ process is the 
Plane Wave Impulse Approximation (PWIA), where one also makes the 
assumption that the proton is ejected from the nucleus without any 
further interaction with the residual nucleus. In nonrelativistic 
PWIA the differential cross section factorizes into two terms, the 
elementary electron-proton cross section, accounting for the 
interaction between the incident electron and the bound proton, and 
the spectral function that accounts for the probability to find a 
proton with given energy and momentum in the nucleus. Although the 
factorization is destroyed when one takes into account the 
distortion of the electron and/or outgoing proton waves, it is 
useful and common practice to analyze the results in terms of a
reduced cross section defined in such a way that would coincide 
with the spectral function if factorization were fulfilled. For 
selected values of the missing energy $E_m$ (i.e., for selected 
single-particle shells) the reduced cross section is given by
\begin{equation}
\rho({\bf p}_m)= \int_{\Delta E_m} dE_m \left[ \sigma^{ep}|
{\bf p}_F|E_F\right] ^{-1} \frac{d^6\sigma}{d E_F d\epsilon_f 
d\Omega_F d\Omega_f} \; ,
\label{reduc}
\end{equation}
with ${\bf p}_m$ the missing momentum, $E_F,|{\bf p}_F|,\Omega_F$ 
($\epsilon_f, \Omega_f$) the outgoing proton (electron) kinematical
variables, and experimentally, the integral is performed over the
interval $\Delta E_m$ that contains the peak of the transition 
under study. The term $\sigma^{ep}$ represents the elementary 
electron-proton cross section. The experimental data of 
$\rho({\bf p}_m)$ are obtained dividing the experimental 
cross section by $\sigma^{ep}_{cc1}$, as given by Eq.(17) of 
Ref.~\cite{deF83}. We therefore use the same expression for 
$\sigma^{ep}$ in our theoretical calculations. In PWIA  
$\rho({\bf p}_m)$ represents the momentum distribution of the 
selected single-particle shell. The spectroscopic factor $S_\alpha$ 
for a given $\alpha$-shell is determined by scaling the theoretical 
predictions for $\rho({\bf p}_m)$ to the experimental data. 

The standard nonrelativistic formalism \cite{Quint} involves the 
{\sc dweepy} program, which is based on an expansion of the one-body 
current operator to second order in the momenta, and can be 
schematically described as follows. The nonrelativistic wave 
functions for the bound and outgoing nucleons are obtained from 
phenomenological potentials of the Woods-Saxon type. The parameters 
of the Woods-Saxon potential for the bound proton are adjusted
for each individual shell. The optical potential for the outgoing 
proton is fitted to elastic proton scattering data. The Coulomb 
distortion of the electron waves is treated in an approximate way.
 
The initial motivation of the fully relativistic formalism was to
incorporate in an exact way the effect of the Coulomb distortion 
on the electron waves~\cite{spect,Jin92}. However, it was soon 
realized~\cite{spect,poten} that this formalism is also more 
adequate than the previous nonrelativistic one in accounting for 
the outgoing proton distortion.

In the relativistic treatment, the nucleons are described by 
solutions of the Dirac equation with Scalar and Vector (S-V) 
potentials. For the bound proton we use the {\sc timora} code 
\cite{HSbook}. The wave function of the outgoing proton is obtained 
by solving the Dirac equation with a S-V optical potential 
\cite{Hama}, fitted to elastic proton scattering data. The complete 
relativistic nucleon current operator with either convention 
\cite{deF83} CC2 or CC1 is used.

Fully relativistic analyses for the quasielastic $(e,e'p)$ reaction 
from the shells $3s_{1/2}$~\cite{spect,Jin92} and $2d_{3/2}$ 
\cite{spect} on $^{208}$Pb have already been made in the low-$p_m$ 
region. The values of the spectroscopic factors obtained with these 
relativistic analyses, $S_\alpha \simeq 0.7$, were much larger than 
the values obtained from previous nonrelativistic analyses 
($S_\alpha \simeq 0.5$) \cite{Quint}. A similar situation was found 
in other doubly closed shell nuclei as $^{40}$Ca. In all cases 
considered, larger spectroscopic factors were obtained with the 
relativistic analyses~\cite{spect,Jin92}. The origin of this 
difference was discussed in detail in Ref.~\cite{poten}. The larger 
values are consistent with theoretical predictions 
\cite{Mahaux,Ma,Pandha} as well as with the spectroscopic factors 
obtained from other methods~\cite{Wagner}. 

In this work we first apply a similar relativistic analysis to the 
low-$p_m$ data for the $1h_{11/2}$, $2d_{5/2}$, and $1g_{7/2}$ 
shells in $^{208}$Pb. The spectroscopic factors resulting from this 
analysis, as well as the ones previously obtained \cite{spect} for 
the $3s_{1/2}$ and $2d_{3/2}$ shells, are then used to calculate the 
reduced cross section in the high-$p_m$ region.

The method used to obtain the spectroscopic factors is as described 
in Ref.~\cite{spect}. For each shell the overall scale factor has 
been obtained by means of an error weighted least-square procedure.
The resulting spectroscopic factors are given in Table 1 for all 
the shells under consideration here. The numbers within parentheses 
correspond to the statistical error derived from the fitting 
procedure. In the two first rows of Table 1 we show the 
spectroscopic factors obtained from the standard nonrelativistic 
analyses of Refs.~\cite{Quint} and~\cite{Irene}, which differ on 
the approximate treatment of the electron Coulomb distortion. 
Note that both nonrelativistic analyses give spectroscopic factors 
that are substantially smaller than ours.

We show in Figures 1 and 2 the reduced cross sections  in the $p_m$ 
range $-100$ MeV $< p_m < 600$ MeV for the five shells in $^{208}$Pb
considered, scaled by the corresponding spectroscopic factors. The 
experimental data in the low-$p_m$ \cite{Quint} and high-$p_m$
\cite{Irene} regions are shown by small and large circles with error 
bars, respectively. 

In the low-$p_m$ region of Figs. 1 and 2 we show our relativistic 
results scaled  by the spectroscopic factors given in the last row 
of Table 1. We can see in Figs. 1 and 2 that the shape of 
$\rho (p_m)$ for each shell agrees very well with data in the 
low-$p_m$ region. This gives confidence on the reliability of these 
spectroscopic factors. As indicated in the figures, these results 
have been obtained using the CC2 current operator. Fits of the 
same quality can be obtained with the CC1 operator \cite{spect}. 
However, the cross sections obtained with the CC1 operator in this 
$p_m$-region are typically $10 \%$ larger than those obtained with 
CC2 and therefore the spectroscopic factors obtained are $10 \%$ 
smaller (see also table 2 of Ref. \cite{spect}). 

In the high-$p_m$ region of figures 1 and 2 we compare with 
experiment our relativistic results  obtained with the current 
operators CC1 and CC2 scaled by the corresponding spectroscopic 
factors. Note that although these two relativistic calculations 
give practically identical results for $\rho (p_m)$ in the 
low-$p_m$ region, they can differ by as much as one order of 
magnitude in the high-$p_m$ region. This difference gives also 
an indication of the theoretical uncertainty that can be expected 
in the high-$p_m$ region even for calculations that fit equally 
well the low-$p_m$ region. Also shown in these figures are the 
results of nonrelativistic calculations  from Ref. \cite{Irene}.

The discontinuity at $p_m=300$ MeV in our theoretical results is 
due to the different kinematics (parallel or perpendicular) in the 
two regions. The main source of this discontinuity can be traced 
back to the electron Coulomb distortion and disappears in the limit 
of plane waves for both electron and proton. A discussion of the 
different effects of electron Coulomb distortion in parallel and
perpendicular kinematics can be found in Ref. \cite{spect}.

One can see from figures 1 and 2 that most of the high-$p_m$ data 
lie between the predictions of the two relativistic calculations, 
while the nonrelativistic calculations underestimate the 
experimental strength. To account for the lack of strength at 
high-$p_m$ in the nonrelativistic calculations, correlations were 
included by Bobeldijk et al. \cite{Irene}, multiplying the bound 
nucleon wave functions by different correlation functions. The 
analysis carried out by these authors showed that the calculations 
including the short-range correlations (SRC) and tensor 
correlations as prescribed by Pandharipande~\cite{Pandha} did not 
modify substantially the mean field predictions. This agrees with 
the conclusion of M\"uther and Dickhoff \cite{Muet}, who find that 
there is no significant increase due to SRC at high momentum and 
low excitation energy compared to the mean field result. On the
contrary, the momentum distributions calculated from quasiparticle 
wave functions given by Mahaux and Sartor~\cite{Mahaux}, and Ma 
and Wambach~\cite{Ma} exhibit an important enhancement in the
high-$p_m$ region and tend to fit better the experimental data. 
This result was considered as an indication of the importance of 
long-range correlations.

Clearly, the relativistic results do not leave much room to claim 
a significant lack of strength in the mean field predictions
at the high momenta and low excitation energies considered here. 
This can be viewed as supporting the remarks in Ref. \cite{Angels} 
in the sense that the relativistic nuclear models could emulate 
the role of correlations. Whether the effect of correlations is 
contained to some extent in the relativistic mean field formalism 
is certainly a point that deserves further study. 

It should be stressed here that we made no attempt to optimize 
agreement with data, and that we use a very simple relativistic 
nuclear model. The bound nucleon wave functions are those obtained 
from the {\sc timora} code without any further adjustment. Taking this 
into account, it is remarkable the good agreement with experiment 
found. It would be interesting to analyze the effect of using 
different relativistic wave functions for both the bound and the 
scattered proton, as well as to study the role of the low components 
of the Dirac wave functions in the high-momentum region. From 
previous studies in Refs.~\cite{poten,Jinnew} we know that the 
effect of the enhancement of the lower  components of the wave 
functions in the relativistic models is very small at low-$p_m$. 
There is work in progress \cite{under} to clarify whether this is 
also the case in the high-$p_m$ region.

Although the data seem to favour the results of the relativistic
calculations, we would like to point out that part of the lack of 
strength in the nonrelativistic result of Ref. \cite{Irene} is due 
to the fact that the normalizing $\sigma^{ep}$ used in the 
theoretical calculations is different from that used in the data.  
In Ref. \cite{Irene} a nonrelativistic approximation 
($\sigma^{ep}_{NR}$) was used in the theoretical calculation of 
$\rho (p_m)$ rather than $\sigma^{ep}_{cc1}$. As can be seen in 
Fig. 3 the nonrelativistic strength in the high-$p_m$ region is 
somewhat increased when $\sigma^{ep}_{cc1}$ is used instead. We 
consider that because $\sigma^{ep}_{cc1}$ has been used in the 
plotted data, the same expression should be used in the theoretical
calculations when comparing to data.

In  Fig. 3 we have also shown the negative missing momentum region. 
This region corresponds to  a similar kinematics as the high-$p_m$ 
region so far discussed except that the polar proton angle is 
different ($p_m>0$ corresponds to $\phi=180^o$, while $p_m<0$ 
corresponds to $\phi=0^o$). Thus, the only difference in the 
cross section is the sign in front of the Longitudinal-Transverse 
(LT) contribution \cite{Pickl}, which is different in each region.
One should keep in mind that if factorization were fulfilled the 
results in both regions should be exactly symmetric. It is 
interesting to observe that the relativistic results are less 
symmetric than the nonrelativistic ones and therefore, the deviation 
between the relativistic and the nonrelativistic results in the 
$p_m<0$ region is enhanced with respect to the one seen in the 
$p_m > 0$ region. It would therefore be interesting to probe the
$p_m < 0$ region experimentally.

In conclusion, we find that compared to the standard nonrelativistic 
results the reduced cross sections obtained with the relativistic 
formalism are quenched in the low-$p_m$ region and enhanced in the 
high-$p_m$ region for the five shells considered. The resulting 
spectroscopic factors are then larger and the profile of the 
momentum distributions agree better with experiment. A clear 
success of the relativistic analysis is the high quality fits to 
the low-$p_m$ data found in each of the orbitals, even though the 
relativistic mean field and nucleon wave functions have not been
adjusted to specific single-particle properties. The high-$p_m$
data are also fairly well accounted for. From our analysis the
same nucleon mean field wave functions and spectroscopic factors
describing the low-$p_m$ data seem to be valid in the high-$p_m$
region discussed here. We would like to emphasize that this
high-$p_m$ region is very sensitive to theoretical models, not 
only to relativistic or nonrelativistic approaches, correlated or 
uncorrelated wave functions, but also to the choice of the 
relativistic nucleon current operator. This choice is of prime
importance since further nonrelativistic approaches depend also
on it. Thus, it is desirable to have more experimental information
in high missing momentum regions. Particularly interesting will be
to explore the $p_m<0$ region that has been found here to depend
more strongly on whether a relativistic or nonrelativistic
approach is used.

We thank I. Bobeldijk for providing us with the experimental data 
and the nonrelativistic single-particle cross sections used in the 
nonrelativistic analyses of Ref.~\cite{Irene} and G. van der 
Steenhoven for useful comments about the NIKHEF-K experiments. 
One of us (JMU) acknowledges support from the EC-program `Human 
Capital and Mobility' under Contract N. CHRX-CT 93-0323.
This work has been partially supported by DGICYT (Spain) under 
Contract 92/0021-C02-01.

\begin{figure}
\caption{Reduced cross sections versus missing momentum for the 
shells $3s_{1/2}$ and $2d_{5/2}$ of $^{208}$Pb. In the low-$p_m$ 
region we show by solid lines the relativistic results scaled 
with the spectroscopic factors of the last row in table 1. Small 
circles with error bars are data from Ref. \protect\cite{Quint}. 
In the high-$p_m$ region we show the relativistic results obtained 
with the currents CC2 (solid lines) and CC1 (long-dashed lines), 
as well as the nonrelativistic results (short-dashed lines) and 
the experimental data from Ref. \protect\cite{Irene}.}
\end{figure}

\vskip 1cm

\begin{figure}
\caption{Same as Fig. 1 but for the shells $1g_{7/2}$, $2d_{3/2}$, 
and $1h_{11/2}$. }
\end{figure}

\vskip 1cm

\begin{figure}
\caption{Reduced cross sections for the $3s_{1/2}$ shell in 
both the positive and negative high-$p_m$ regions. Circles
with error bars are data from Ref. \protect\cite{Irene}. We show
relativistic calculations obtained with the currents CC2 (solid 
lines) and CC1 (long-dashed lines), and nonrelativistic calculations 
normalized with $\sigma_{cc1}^{ep}$ (dashed lines) and 
$\sigma_{NR}^{ep}$ (short-dashed lines).}
\end{figure}
\newpage
\mediumtext

\begin{table}
\caption{Spectroscopic factors deduced from  the relativistic and
nonrelativistic analyses of the low-$p_m$ data in the reaction
$^{208}$Pb$(e,e'p)^{207}$Tl. The numbers within parentheses
indicate the statistical error derived from the fit.
\label{spectf} }
\vskip .4cm
\begin{tabular}{lccccc}
  & $3s_{1/2}$ & $2d_{3/2}$ & $1h_{11/2}$ & $2d_{5/2}$ & $1g_{7/2}$ 
  \\ \tableline
Nonrel. (Ref. 
\protect\cite{Quint})  & .50  & .53    & .42    & .44    & .19  \\ 
Nonrel. (Ref. 
\protect\cite{Irene})  & .55  & .57    & .58    & .54    & .26  \\
Rel. (this work and Ref. 
\protect\cite{spect})  & .70(5) & .73(5) & .64(4) & .60(5) & .30(4) 
\end{tabular}
\end{table}

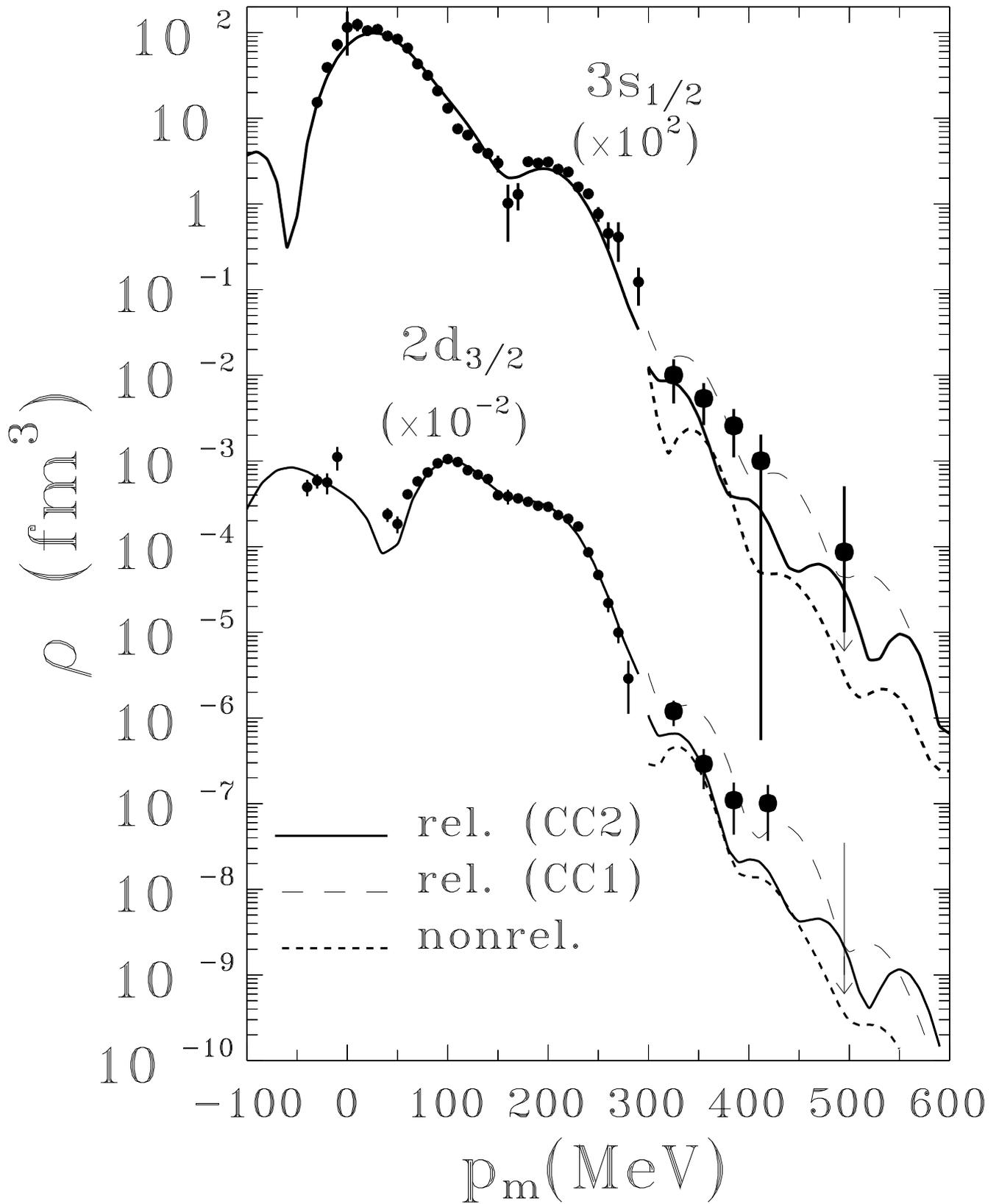
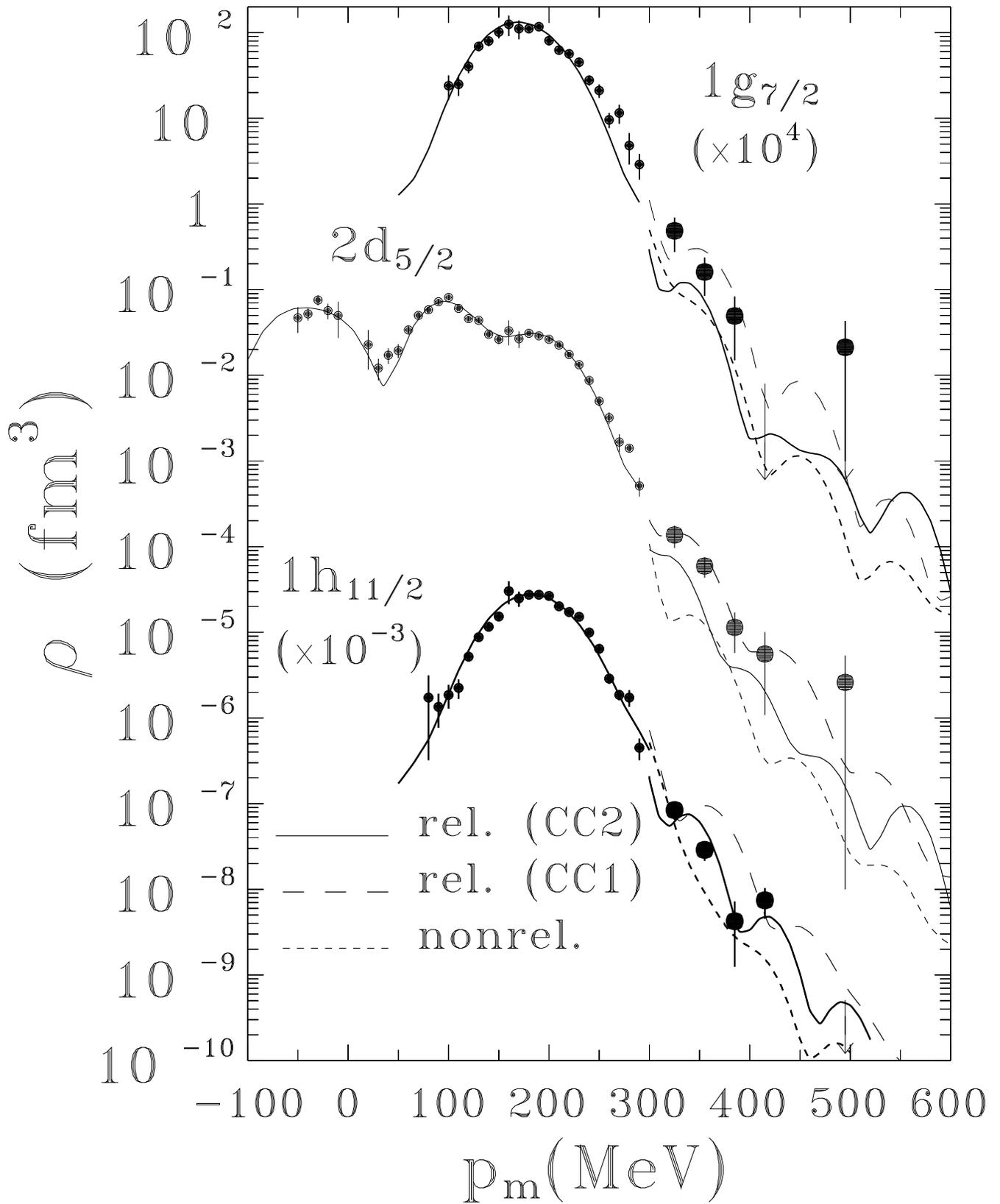
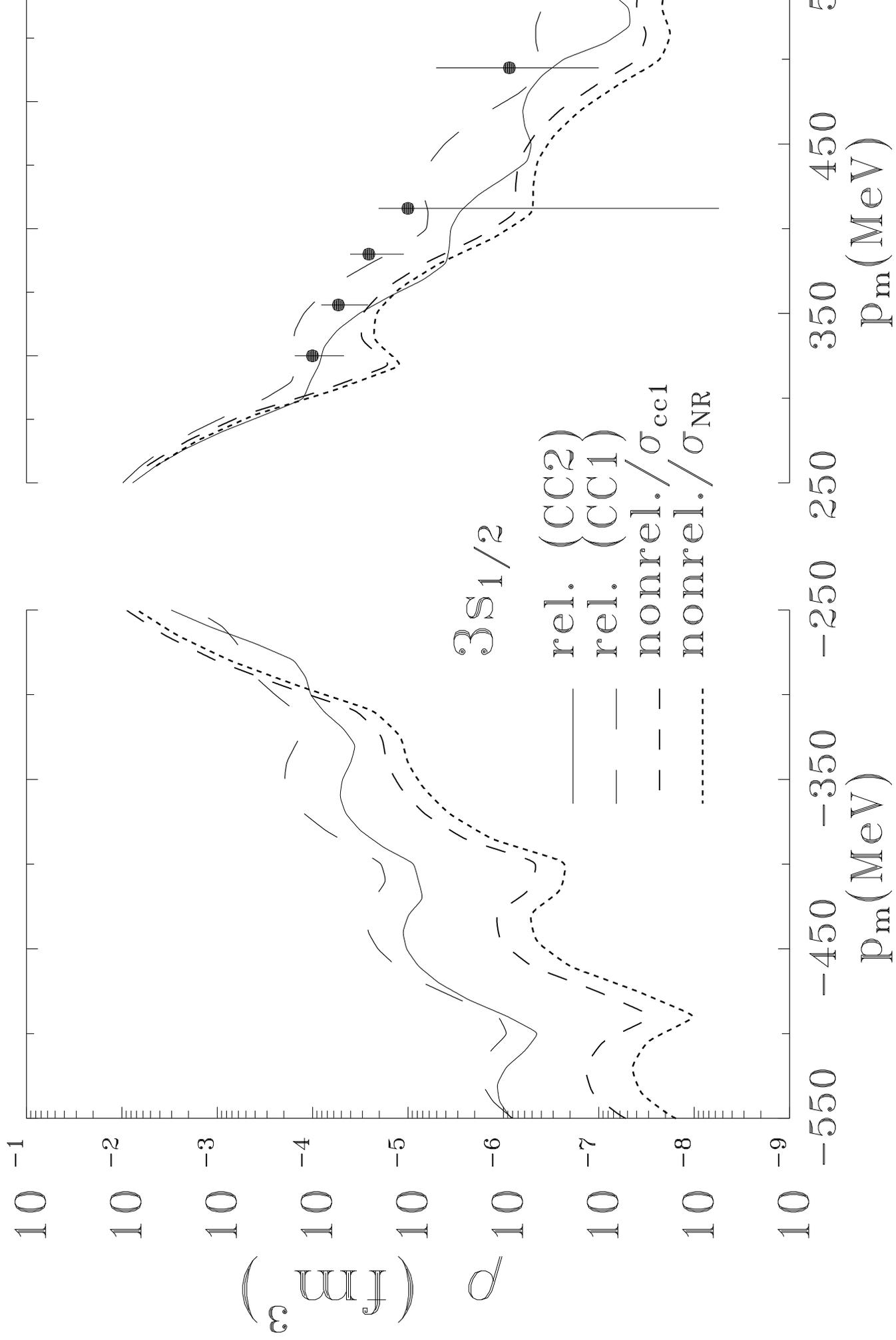
\end{document}